\documentstyle[twoside,fleqn,bezier]{article}

\newcommand{\bls}[1]{\renewcommand{\baselinestretch}{#1}}
\def\noi{\noindent}

\makeatletter

\renewcommand{\section}{\@startsection{section}{1}{0pt}%
        {-3.5ex plus -1ex minus -.2ex}{2.3ex plus .2ex}%
        {\large\bf\protect\raggedright}}

\renewcommand{\subsection}{\@startsection{subsection}{2}{0pt}%
        {-3ex plus -1ex minus -.2ex}{1.4ex plus .2ex}%
        {\normalsize\bf\protect\raggedright}}

\renewcommand{\@oddhead}{\raisebox{0pt}[\headheight][0pt]{%
   \vbox{\hbox to\textwidth{\rightmark \hfil \rm \thepage \strut}\hrule}}}
\renewcommand{\@evenhead}{\raisebox{0pt}[\headheight][0pt]{%
   \vbox{\hbox to\textwidth{\thepage \hfil \leftmark \strut}\hrule}}}
\newcommand{\heads}[2]{\markboth{\protect\small\it #1}{\protect\small\it #2}}
\newcommand{\Acknow}[1]{\subsection*{Acknowledgement} #1}

\makeatother

\newcommand{\Title}[1]{\noindent {\Large #1} \\}
\newcommand{\Author}[2]{\noindent{\large\bf #1}\\[2ex]\noindent{\it #2}\\}
\newcommand{\Abstract}[1]{\vskip 2mm \begin{center}
     \parbox{16.4cm}{\small\noindent #1} \end{center}\bigskip}
\newcommand{\foom}[1]{\protect\footnotemark[#1]}
\newcommand{\foox}[2]{\footnotetext[#1]{#2}}
\newcommand{\email}[2]{\footnotetext[#1]{e-mail: #2}}

\newcommand{\sect}[1]{Sec.\,#1}
\newcommand{\ssect}[1]{Subsec.\,#1}

\def\nq{\hspace{-1em}}
\def\nqq{\hspace{-2em}}
\def\nhq{\hspace{-0.5em}}

\def\cm{\hspace{1cm}}
\def\inch{\hspace{1in}}

\def\eq{Eq.\,}
\def\eqs{Eqs.\,}
\def\beq{\begin{equation}}
\def\eeq{\end{equation}}
\def\bear{\begin{eqnarray}}
\def\al{&\nhq}
\def\lal{&&\nqq {}}               
\def\bearr{\begin{eqnarray} \lal}
\def\ear{\end{eqnarray}}
\def\earn{\nonumber \end{eqnarray}}
\def\dst{\displaystyle}
\def\tst{\textstyle}
\newcommand{\fracd}[2]{{\dst\frac{#1}{#2}}}

\def\nn{\nonumber\\ {}}

\def\nnn{\nonumber\\ \lal }

\def\yy{\\[5pt]}

\def\eql{\al =\al}

\def\eqdef{\stackrel{\rm def}{=}}
\def\e{{\,\rm e}}
\def\d{\partial}

\def\sign{{\,\rm sign\,}}

\def\const{{\rm const}}

\def\half{{\tst\frac{1}{2}}}

\def\DAL{\raisebox{-1.6pt}{\large $\Box$}\,}
\newcommand{\vars}[1]{\left\{\begin{array}{ll}#1\end{array}\right.}

\def\eps{\varepsilon}
\def\umx{u_{\max}}
\def\sph{spherically symmetric\ }

\def\bh{black hole}
\def\bhs{black holes}

\def\df{\delta\varphi}
\def\da{\delta\alpha}
\def\db{\delta\beta}
\def\dg{\delta\gamma}
\def\pn{perturbation}
\def\pns{perturbations}
\def\fig{Fig.\,}

\newcommand{\Picture}[3]{
	\begin{figure} \centering \unitlength=1mm
	\begin{picture}(82.5,#1)
		\put(0,0){\line(0,1){#1}}        
		\put(0,0){\line(1,0){82.5}}              
		\put(82.5,0){\line(0,1){#1}}             
		\put(0,#1){\line(1,0){82.5}}
	\put(0,0){#2}                       \end{picture}
        \caption{\protect\small #3}  \medskip \hrule \end{figure} }

\newcommand{\WPicture}[3]{
	\begin{figure*}[!t] \centering \unitlength=1mm
	\begin{picture}(174,#1)
		\put(0,0){\line(0,1){#1}}        
		\put(0,0){\line(1,0){174}}               
		\put(174,0){\line(0,1){#1}}              
		\put(0,#1){\line(1,0){174}}
	\put(0,0){#2}                        \end{picture}
        \caption{\protect\small #3}  \medskip \hrule \end{figure*} }

\heads{K.A. Bronnikov, G. Cl\'ement, C.P. Constantinidis and J.C. Fabris}
{Cold Scalar-Tensor Black Holes: Causal Structure, Geodesics, Stability}

\bls{1.05}

\begin{document}
\thispagestyle{empty}
\twocolumn[
\rightline{\bf gr-qc/9804064}

\vspace*{25mm}

\Title{COLD SCALAR-TENSOR BLACK HOLES:           \yy
       CAUSAL STRUCTURE, GEODESICS, STABILITY}

\Author{K.A. Bronnikov\foom 1,
        G. Cl\'ement\foom 2,
        C.P. Constantinidis\foom 3
        and J.C. Fabris\foom 4}
{Departamento de F\'{\i}sica, Universidade Federal do Esp\'{\i}rito Santo,
Vit\'oria, Esp\'{\i}rito Santo, Brazil}

\Abstract
{We study the structure and stability of the recently discussed \sph
Brans-Dicke black-hole type solutions with an infinite horizon area and zero
Hawking temperature,
existing for negative values of the coupling constant $\omega$. These
solutions split into two classes: B1, whose horizon is reached by an
infalling particle in a finite proper time, and B2, for which this proper
time is infinite. Class B1 metrics are shown to be extendable beyond the
horizon only for discrete values of mass and scalar charge, depending on two
integers $m$ and $n$. In the case of even $m-n$ the space-time is globally
regular; for odd $m$ the metric changes its signature as the horizon is
crossed. Nevertheless, the Lorentzian nature of the space-time is
preserved, and geodesics are smoothly continued across the horizon, but,
when crossing the horizon, for odd $m$ timelike geodesics become spacelike
and {\sl vice versa}. Problems with causality, arising in some cases, are
discussed. Tidal forces are shown to grow infinitely near type B1 horizons.
All vacuum static, \sph solutions of the Brans-Dicke
theory with $\omega<-3/2$ are found to be linearly stable against \sph \pns.
This result extends to the generic case of the Bergmann-Wagoner class of
scalar-tensor theories of gravity with the coupling function
$\omega(\phi) < -3/2$.  }

]  
\foox 1 {e-mail: kb@goga.mainet.msk.su;
	permanent address: Centre for Gravitation and Fundamental Metrology,
    	VNIIMS, 3-1 M. Ulyanovoy St., Moscow 117313, Russia.}

\section{Introduction}

    In the recent years there has been a renewed interest in scalar-tensor
    theories (STT) of gravity as viable alternatives to general relativity
    (GR), mostly in connection with their possible role in the early
    Universe:
    they provide power-law instead of exponential inflation, leading to
    more plausible perturbation spectra \cite{spectra}.
    Another aspect of interest in STT is the possible existence of
    black holes (BHs) different from those well-known in GR.
\foox 2 {e-mail: gecl@ccr.jussieu.fr; permanent address:
     Laboratoire de Gravitation et Cosmologie Relativistes, Universit\'e
     Pierre et Marie Curie, CNRS/URA769, Tour 22-12, Bo\^{\i}te 142, 4 Place
     Jussieu, 75252 Paris Cedex 05, France.}
\email 3 {clisthen@cce.ufes.br}
\email 4 {fabris@cce.ufes.br}
\setcounter{footnote}{4}
    Thus, Campanelli and Lousto \cite{lousto} pointed out among the
    static, \sph solutions of the Brans-Dicke (BD) theory
    a subfamily possessing all BH properties, but (i) existing only for
    negative values of the coupling constant $\omega$ and (ii) with
    horizons of infinite area. These
    authors argued that large negative $\omega$ are compatible with modern
    observations and that such BHs may be of astrophysical relevance%
\footnote
{For brevity, we call BHs with infinite horizon areas type B BHs \cite{we},
to distingish them from the conventional ones,
with finite horizon areas, to be called type A.}.
    Indeed, the post-Newtonian parameters, determining the observational
    behaviour of STT in the weak field limit, depend crucially on the
    absolute value of $\omega$ rather than its sign \cite{73,will}.

    In Ref.\,\cite{we} it was shown, in the framework of a general
    (Bergmann-Wagoner) class of STT, that nontrivial BH solutions can exist
    for the coupling function $\omega(\phi)+3/2 <0$, and that only in
    exceptional cases these BHs have a finite horizon area. In the BD theory
    ($\omega = \const$) such BHs were indicated explicitly; they have
    infinite horizon areas and zero Hawking temperature (``cold BHs''), thus
    confirming the conclusions of \cite{lousto}.  These BHs in turn split
    into two subclasses: B1, where horizons are attained by infalling
    particles in a finite proper time $\tau$, and B2, for which $\tau$
    is infinite. These results are briefly presented in Sections 2 and 3.

    The static region of a type B2 BH is geodesically complete since its
    horizon is infinitely remote and actually forms a second spatial
    asymptotic.  For type B1 BHs the global picture is more complex and is
    discussed here in some detail (\sect 4). It turns out that the horizon
    is generically singular due to violation of analyticity, despite the
    vanishing curvature invariants. Only a discrete set of B1-solutions,
    parametrized by two integers $m$ and $n$, admits a Kruskal-like
    extension, and, depending on their parity, four different global
    structures are distinguished. Two of them, where $m-n$ is even, are
    globally regular, in two others the region beyond the horizon contains a
    spacelike singularity.

    \ssect {4.4} describes the behaviour of geodesics for different cases of
    BD BH metrics. It turns out that for odd $m$,
    when crossing the horizon, timelike geodesics become spacelike and {\it
    vice versa\/}, leading to problems with causality.
    Thus, in a family of BHs there appear closed timelike curves, whose
    existence may be avoided by assuming a ``helicoidal" analytic extension
    of the space-time manifold. Moreover, it is shown by considering the
    geodesic deviation equations that near a horizon with an infinite area
    all extended bodies are destroyed (stretched apart) by unbounded tidal
    forces.

    \sect 5 discusses the stability of STT solutions under \sph \pns. In
    this case the only dynamical degree of freedom is the scalar field, and
    the \pn\ analysis reduces to a single wave equation whose radial
    part determines the system behaviour. Under reasonable boundary
    conditions, it turns out that the BD solutions with $\omega<-3/2$ are
    linearly stable, and this result extends to similar solutions of the
    general STT provided the scalar field does not create new singularities
    in the static domain.  For the case $\omega > -3/2$ the stability
    conclusion depends on the boundary condition at a naked singularity,
    which is hard to formulate unambiguously.

    We can conclude that, in STT with the anomalous sign of the scalar
    field energy, vacuum \sph solutions generically describe stable BH-like
    configurations. Some of them, satisfying a ``quantization" condition,
    are globally regular and have peculiar global structures. We also
    conclude that, due to infinite tidal forces, a horizon with an infinite
    area converts any infalling body into true elementary
    (pointlike) particles, which afterwards become tachyons.

    A short preliminary account of this work has been given in \cite{PLA}.
    In our next paper we intend to discuss similar solutions with an
    electric charge.


\section {Field equations}  

     The Lagrangian of the general (Bergmann-Wagoner) class of
     STT of gravity in four dimensions is
\beq                                               \label{L1}
     L = \sqrt{-g}\biggr[\phi R
           + \frac{\omega(\phi)}{\phi}\phi_{;\rho}\phi^{;\rho}
	                     + L_{\rm m}\biggl]
\eeq
     where $\omega(\phi)$ is an arbitrary function of the scalar field
     $\phi$ and $L_{\rm m}$ is the Lagrangian of non-gravitational matter.
     This formulation (the so-called {\it Jordan conformal
	frame\/}) is commonly considered to be fundamental since just in this
	frame the matter energy-momentum tensor $T^\mu_\nu$ obeys the
	conventional conservation law $\nabla_{\alpha}T^\alpha_\mu =0$, giving
	the usual equations of motion (the so-called atomic system of
	measurements). In particular, free particles move along geodesics of
	the Jordan-frame metric. Therefore, in what folows we discuss the
	geometry, causal structure and geodesics in the Jordan frame.

	We consider only scalar-vacuum configurations and put
	$L_{\rm m}=0$.

	The field equations are easier to deal with in
     the {\it Einstein conformal frame\/}, where the transformed scalar
	field $\varphi$ is minimally coupled to gravity. Namely, the
	conformal mapping
     $g_{\mu\nu} = \phi^{-1}\bar g_{\mu\nu}$ transforms \eq (\ref{L1})
	(up to a total divergence) to
\bearr                                                  \label{L2}
     L = \sqrt{-\bar g}\biggr(\bar R + \eps
         {\bar g}^{\alpha\beta}\varphi_{;\alpha}\varphi_{;\beta}\biggl),
	\\  \lal
             \eps = \sign (\omega + 3/2),                  \label{eps}
	\cm
   \frac{d\varphi}{d\phi} = \biggl|\frac{\omega + 3/2}{\phi^2}\biggr|^{1/2}.
\ear
     The field equations are
\beq                                                         \label{eq1}
     R_{\mu\nu} = -\eps \varphi_\mu \varphi_\nu, \cm
     \nabla^\alpha \nabla_\alpha\varphi =0
\eeq
     where we have suppressed the bars marking the Einstein frame.
	The value $\eps=+1$ corresponds to normal STT, with positive energy
	density in the Einstein frame; the choice $\eps=-1$ is anomalous. The
	BD theory corresponds to the special case $\omega = \const$, so
	that
\beq
	\phi = \exp (\varphi/\sqrt{|\omega+3/2|}).      \label{phi-bd}
\eeq

     Let us consider a \sph field system, with the metric
\bearr                                                 \label{met1}
     ds_{\rm E}^2 = \e^{2\gamma}dt^2 - \e^{2\alpha}du^2
                  - \e^{2\beta}d\Omega^2,  \nnn
	\cm d\Omega^2 =
			   d\theta^2 + \sin^2 \theta d\Phi^2,
\ear
     where E stands for the Einstein frame, $u$ is the radial
     coordinate, $\alpha$, $\beta$, $\gamma$ and the field $\varphi$ are
	functions of $u$ and $t$.
     Preserving only linear terms with respect to
     time derivatives, we can write \eqs (\ref{eq1}) in the following form:
\bear
     \e^{2\alpha}R^0_0 \eql                                   \label{00}
     \e^{2\alpha-2\gamma}(\ddot\alpha + 2\ddot\beta)  \nnn
           -[ \gamma'' +\gamma'(\gamma'-\alpha'+2\beta')] =0; \\
     \e^{2\alpha}R^1_1 \eql
     \e^{2\alpha-2\gamma}\ddot\alpha                  \nnn    \label{11}
 \nqq\nqq\nq  -[\gamma''+2\beta'' +\gamma'{}^2+2\beta'{}^2
		    -\alpha'(\gamma'+2\beta')]  = \half\eps\varphi'{}^2;\\
     \e^{2\alpha}R^2_2 \eql
     \e^{2\alpha-2\beta}                                       \label{22}
          +\e^{2\alpha-2\gamma}\ddot\beta              \nnn
	          -[\beta''+\beta'(\gamma'-\alpha'+2\beta')] =0;\\
     R_{01}\eql
     2[\dot\beta' + \dot{\beta}\beta'                          \label{01}
                 -\dot{\alpha}\beta'-\dot{\beta}\gamma']
				  = -\eps \dot{\varphi}\varphi'; \\
     \sqrt{-g}\DAL \varphi\eql                                 \label{ephi}
	     \e^{-\gamma+\alpha+2\beta}\ddot\varphi
	        -\bigl(\e^{\gamma-\alpha+2\beta} \varphi'\bigr)' =0
\ear
     where dots and primes denote, respectively, $\d/\d t$ and $\d/\d u$.

	Up to the end of \sect 4 we will
	restrict ourselves to static configurations.


\section{Black holes in scalar-tensor theories}           

     The general static, \sph scalar-vacu\-um solution of the
     theory (\ref{L1}) is given by \cite{73,k1}
\bearr                                                     \label{ds1}
     ds_{\rm J}^2 = \frac{1}{\phi}ds_{\rm E}^2 \nnn
     = \frac{1}{\phi}
     \biggl\{\e^{-2bu}dt^2 - \frac{\e^{2bu}}{s^2(k,u)}
       \biggr[\frac{du^2}{s^2(k,u)} + d\Omega^2\biggl]
	     \biggr\},
		\\  \lal                                       \label{phi1}
     \varphi = Cu + \varphi_0, \cm C, \varphi_0 =\const,
\ear
     where J denotes the Jordan frame,
     $u$ is the harmonic radial coordinate in the static space-time
	in the Einstein frame,
     such that $\alpha(u) = 2\beta(u) + \gamma(u)$, and the function
	$s(k,u)$ is
\beq                                                \label{def-s}
     s(k,u) \eqdef \vars     {
                    k^{-1}\sinh ku,  \ & k > 0, \\
                                  u, \ & k = 0, \\
                    k^{-1}\sin ku,   \ & k < 0,     }
\eeq
     The constants $b$, $k$ and $C$ (the scalar charge) are related by
\beq                                                      \label{r1}
            2k^2\sign k = 2b^2 + \eps C^2.
\eeq
     The range of $u$ is $0 < u < \umx$, where $u=0$ corresponds to spatial
	infinity, while $\umx$ may be finite or infinite depending on $k$ and
	the behaviour of $\phi(\varphi)$ described by (\ref{eps}). In normal
	STT ($\eps=+1$), by (\ref{r1}), we have only $k > 0$, while in
	anomalous STT $k$ can have either sign.

     According to the previous studies \cite{73,k1}, all these solutions in
     normal STT have naked singularities, up to rare exceptions
     when the sphere $u=\infty$ is regular and admits an extension of the
	static coordinate chart. An example is a conformal scalar field in
	GR viewed as a special case of STT, leading to BHs with scalar charge
	\cite{70,bek}. Even when it is the case, such configurations are
	unstable due to blowing-up of the effective gravitational coupling
	\cite{78}.

     In anomalous STT ($\eps=-1$) the solution behaviour is more diverse and
	the following cases without naked singularities can be found:

\medskip\noi
     {\bf 1.} $k > 0$.
     Possible event horizons have an infinite area (type B BHs),
     i.e. $g_{22}\to\infty$ as $r\to 2k$.
     In BD theory, after the coordinate transformation
\beq
     \e^{-2ku} = 1 - 2k/r \equiv P(r)                    \label{def-P}
\eeq
	the solution takes the form
\bear                                                       \label{ds+}
     ds_{\rm J}^2 \eql P^{-\xi} ds^2_{\rm E} \nn
                  \eql P^{-\xi}\Bigl(P^{a }dt^2 - P^{-a }dr^2
                                 - P^{1 - a }r^2 d\Omega^2 \Bigl),\nn
	\phi \eql P^\xi
\ear
     with the constants related by
\beq
     (2\omega+3) \xi^2 = 1-a^2, \cm a=b/k.               \label{int+}
\eeq
     The allowed range of $a$ and $\xi$, providing
     a horizon without a curvature singularity at $r=2k$, is
\beq
	a >1, \cm\cm     a  > \xi \geq 2- a.                 \label{range+}
\eeq
	(\eqs (\ref{ds+}),(\ref{int+}) are valid for $\omega>-3/2$ as well, but
	then $a<1$, leading to a naked singularity.)

     For $\xi < 1$ particles can arrive at the horizon in a finite proper
	time and may eventually (if geodesics can be extended)
	cross it, entering the BH interior
     (type B1 BHs \cite{we}).
     When $\xi\geq 1$, the sphere $r=2k$ is infinitely far and it takes
	an infinite proper time for a particle to reach it. Since in the
	same limit $g_{22}\to \infty$, this configuration (a type
	B2 BH \cite{we}) resembles a wormhole.

\medskip\noi
     {\bf 2.} $k = 0$.
     Just as for $k>0$, in a general STT, only type B \bhs\ are possible
     \cite{we}), with $g_{22}\to\infty$ as $u\to \infty$.
     In particular, the BD solution is
\bearr
     ds^2 = \e^{-su}\biggr[\e^{-2bu}dt^2 - \frac{\e^{2b u}}{u^2}\biggr(
	\frac{du^2}{u^2} + d\Omega^2\biggl)\biggl], \nnn
		  \phi = \e^{su}, \cm
		      s^2(\omega + 3/2) = -2b^2.               \label{ds0}
\ear
     The allowed range of the integration constants is
     $b > 0, \quad 2b > s > -2b$.
    This range is again divided into two halves: for $s>0$ we
    deal with a type B1 BH, for $s<0$ with that of type B2
    ($s=0$ is excluded since it leads to GR).

\medskip\noi
    {\bf 3.} $k < 0$.
    In the general STT the metric (\ref{ds1}) describes a wormhole, with
    two flat asymptotics at $u=0$ and $u=\pi/|k|$, provided $\phi$ is
    regular between them. In exceptional cases
    the sphere $\umx = \pi/|k|$ may be an event horizon, namely, if
    $\phi \sim 1/\Delta u^2$, $\Delta u \equiv |u - \umx| $. In this case
    it has a finite area (a type A \bh) and
\beq                                                           \label{bh-}
    \omega(\phi) + 3/2 \to  -0 \qquad {\rm as} \qquad  u\to\umx.
\eeq
    The behaviour of $g_{00}$ and $g_{11}$ near the horizon is then similar
    to that in the extreme Reissner-Nordstr\"om solution.

    In the BD theory we have only the wormhole solution
\bearr
	ds^2 = \e^{-su}\biggr[\e^{-2bu}dt^2                       \label{ds-}
	     - \frac{k^2\e^{2bu}}{\sin^2{ku}}\biggr(
	\frac{k^2du^2}{\sin^2{ku}} + d\Omega^2\biggl)\biggl],
\nnn
        s^2(\omega + 3/2) = - k^2 - 2b^2.
\ear
    with masses of different signs at the two asymptotics.
\medskip\noi

    For all the BH solutions mentioned, the Hawking temperature is zero.
    Their infinite horizon areas may mean that their entropy is also
    infinite; however, a straightforward application of the
    proportionality relation from GR between entropy and horizon area,
    is here hardly justified; a calculation of BH entropy is a separate
    problem, discussed in a large number of recent works from various
    standpoints.


\section{Analytic extension and causal structure of Brans-Dicke
         black holes}                                  

\subsection {Extension}                                

    Let us discuss possible Kruskal-like extensions of type B1
    BH metrics (\ref{ds+}) ($k>0$) and (\ref{ds0}) ($k=0$) of the BD theory.

    For (\ref{ds+}), with $a > 1 > \xi > 2-a$,
    we introduce, as usual, the null coordinates $v$ and $w$:
\beq
    v = t + x, \qquad w = t - x, \qquad
                                  x \eqdef \int P^{-a}dr   \label{vw}
\eeq
    where $x \to \infty$ as $r \to \infty$ and $x \to -\infty$
    as $r \to 2k$. The asymptotic behaviour of
    $x$ as $r \to 2k$ ($P \to 0$) is
    $x \propto - P^{1-a}$, and in a finite
    neighbourhood of the horizon $P=0$ one can write
\beq                                                        \label{x}
    x \equiv \half(v-w) =-\half P^{1-a}f(P)\,,
\eeq
    where $f(P)$ is an analytic function of $P$, with $f(0)=4k/(a-1)$:
\beq                                                 \label{f}
	f(P) = -4k\sum_{q=0}^{\infty} \frac{q+1}{q-a+1}P^q.
\eeq
    Then, let us define new null coordinates $V<0$ and $W>0$
    related to $v$ and $w$ by
\beq                                                        \label{VW}
       - v = (-V)^{-n-1}, \quad\  w = W^{-n-1}, \quad\  n=\const.
\eeq
    The mixed coordinate patch $(V,w)$ is defined for $v<0$ ($t<-x$) and
    covers the whole past horizon $v=-\infty$. Similarly, the patch $(v,W)$
    is defined for $w>0$ ($t>x$) and covers the whole future horizon
    $w=+\infty$. So these patches can be used to extend the metric through
    one or the other horizon.

    Consider the future horizon. As is easily verified, a finite value
    of the metric coefficient $g_{vW}$ at $W=0$ is achieved if we take
    $n+1 = (a-1)/(1-\xi)$, which is positive for $a>1>\xi$.
    In a finite neighbourhood of the horizon,
    the metric (\ref{ds+}) can be written in
    the coordinates $(v,W)$ as follows:
\bear                                                        \label{ext}
    ds^2 \eql P^{a-\xi} dv\, dw - P^{1-a-\xi}r^2 d\Omega^2
\nn
    \eql -(n+1)f^{\frac{n+2}{n+1}}
              \cdot(1-vW^{n+1})^{-\frac{n+2}{n+1}}dv\,dW
\nnn
     -\fracd{4k^2}{(1-P)^2}f^{-\frac m{n+1}}
                    \cdot(1-vW^{n+1})^{\frac m{n+1}} W^{-m} d\Omega^2\nnn
\ear
    where
\beq
    m = (a-1+\xi)/(1-\xi).                  \label{def-m}
\eeq
    The metric (\ref{ext}) can be extended at fixed $v$ from $W>0$ to $W<0$
    only if the numbers $n+1$ and $m$ are both integers (since otherwise
    the fractional powers of negative numbers violate the analyticity in
    the immediate neighbourhood of the horizon).  This leads to a discrete
    set of values of the integration constants $a$ and $\xi$:
\beq
          a = \frac{m+1}{m-n}, \cm    \xi = \frac{m-n-1}{m-n}.   \label{qu}
\eeq
    where, according to the regularity conditions (\ref{range+}),
    $m > n \geq 0$. Excluding the Schwarzschild case $m=n+1$ ($\xi =0$), which
    corresponds from (\ref{ds-}) to $a=1$ ($m=0$), we see
    that regular BD BHs correspond to integers $m$ and $n$ such that
\beq
    m-2 \ge n \ge 0.                                            \label{mn}
\eeq

    An extension through the past horizon can be performed in the
    coordinates $(V,w)$ in a similar way and with the same results.

    It follows that, although the curvature scalars
    vanish on the Killing horizon $P=0$, the metric cannot
    be extended beyond it unless the constants $a$ and $\xi$
    obey the ``quantization condition'' (\ref{qu}),
    and is generically singular. The Killing horizon, which
    is at a finite affine distance, is part of the boundary of the
    space-time, i.e. geodesics and other possible trajectories terminate
    there. Similar properties were obtained in a (2+1)--dimensional model
    with exact power--law metric functions \cite{sigma} and in the  case of
    black $p$--branes \cite{GHT}.

\Picture{60}
{\nq \input bh0.pic}
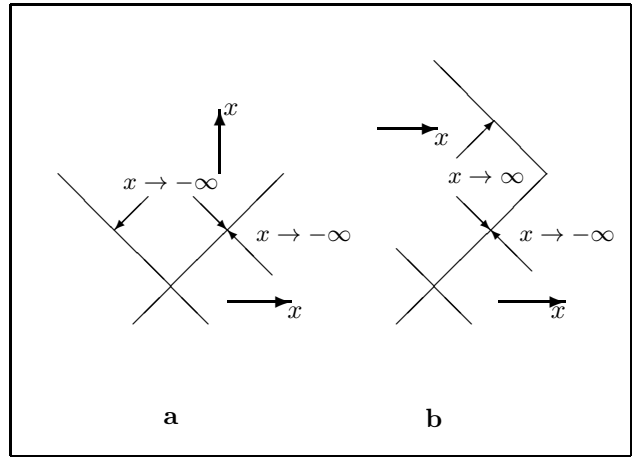
{Extensions through the future horizon for odd (a) and even (b) values of
 $n$. Thick vertical and horizontal arrows show the growth direction of the
 $x$ coordinate in the corresponding regions.}

    We have thus obtained a discrete family of BH solutions whose parameters
    depend on the two integers $m$ and $n$. The corresponding parameters
    describing the asymptotic form of the solution, the active gravitational
    mass $M$ and the scalar charge $S$, defined by
\[
    \nhq  	g_{00}= 1 -\frac{2GM}{r} +o\biggl(\frac 1r\biggr),
	\qquad     \phi = 1 + \frac Sr + o\biggl(\frac 1r\biggr),
\]
    where $G$ is the Newtonian gravitational constant,
    are ``quantized" according to the relations
\beq
    GM = k\frac{n+2}{m-n}, \cm S = -2k\frac{m-n-1}{m-n}.      \label{mass}
\eeq
    The constant $k$, specifying the length scale of the solution,
    remains arbitrary. On the other hand, the coupling constant $\omega$
    takes, according to (\ref{int+}), discrete values:
\beq
	|2\omega+3| = \frac{(2m-n+1)(n+1)}{(m-n-1)^2}.          \label{om-mn}
\eeq

The $k = 0$ solution (\ref{ds0}) of the BD theory also has a Killing
horizon ($u \to \infty$) at finite geodesic distance if $s > 0$. However,
this space-time does not admit a Kruskal--like
extension and so is singular. The reason is that in this case the relation
giving the tortoise--like coordinate $x$,
\beq                                                         \label{x0}
x = \int\frac{{\rm e}^{2bu}}{u^2}\,du = \frac{{\rm e}^{2bu}}{2bu^2}F(u)
\eeq
(where $F(u)$ is some function such that $F(\infty) = 1$) cannot be
inverted near $u = \infty$ to obtain $u$ as an analytic function of $x$.

\subsection {Geometry and causal structure}            

    To study the geometry beyond the horizons of the metric (\ref{ds+}), or
    (\ref{ext}), let us define the new radial coordinate $\rho$ by
\beq                                                         \label{rho}
    P \equiv {\rm e}^{-2ku} \equiv 1 - \frac{2k}{r} \equiv \rho^{m-n}.
\eeq
    The resulting solution (\ref{ds+}), defined in the
    static region I ($P>0$, $\rho > 0$), is
\bear                                                \label{global}
    ds^2 \eql \rho^{n+2}\,dt^2 -
              \frac{4k^2(m-n)^2}{(1-P)^4}\,\rho^{-n-2}\,d\rho^2 \nnn
        - \frac{4k^2}{(1-P)^2}\,\rho^{-m}\,d\Omega^2,
                                                       \qquad
    \phi = \rho^{m-n-1}.
\ear
    By (\ref{x}), $\rho$ is related to the mixed null coordinates
    $(v,W)$ by
\bearr                                                      \label{rhoW}
      \rho (v,W) = W\,[f(P)]^{1/(n+1)}[1-vW^{n+1}]^{-1/(n+1)}.   \nnn
\ear
    This relation and a similar one giving $\rho (V,w)$
    show that when the future (past) horizon
    is crossed, $\rho$ varies smoothly, changing its sign with $W$ ($V$).
    For $\rho < 0$ the metric (\ref{global}) describes
    the space-time regions beyond the horizons.
 \WPicture{79}
     {\qquad      \input bh1.pic   \cm
      \qquad      \input bh2.pic}
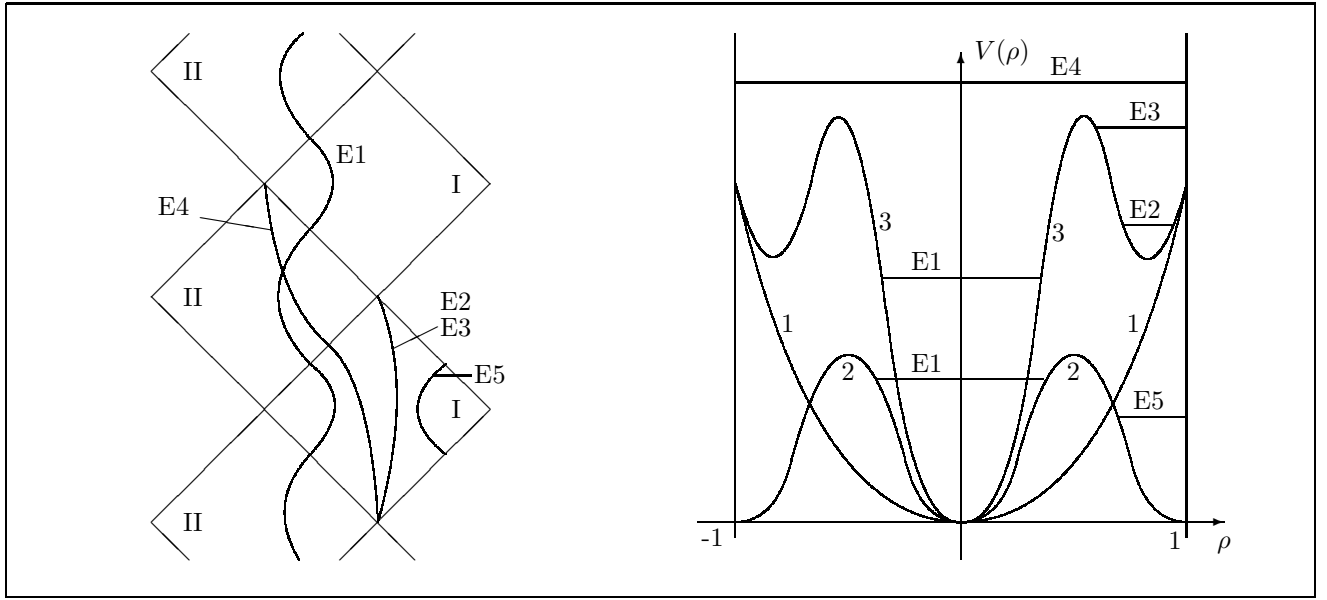
 {The Penrose diagram and the effective potential for geodesics for a BH
 with $m$ and $n$ both even. Curve 1 is $V_\eta(\rho)$, curve 2 is
 $V_L(\rho)$ and curve 3 is $V(\rho)$ for a nonradial timelike geodesic. The
 curves and ``energy levels" E1--E5 correspond to different kinds of
 geodesics as described in \ssect 4.4, item 1a.}

    To construct the corresponding Penrose diagrams, it is helpful to notice
    that by (\ref{x}) the radial coordinate $x$ is related to $\rho$ by
\beq
	x= -\half \rho^{-n-1} f(P),                      \label{x-rho}
\eeq
    so that for odd $n$ the horizon as seen from region II ($\rho <0$) also
    corresponds to $x\to -\infty$. On the other hand, in the 2-dimensional
    metric $ds_2^2 = \rho^{n+2}(dt^2 - dx^2)$, for $\rho <0$ the coordinate
    $x$ is timelike, hence in region II beyond the future horizon (with
    respect to the original region I) $x\to -\infty$ means ``down". A new
    horizon for region II joins the picture at point O --- see \fig 1(a).
    For even $n$, when $x$ in region II remains a spatial coordinate
    and the coordinate (\ref{x-rho}) changes its sign when crossing the
    horizon, a new horizon joins the old one at the future infinity point of
    region I --- \fig 1(b). These considerations are easily verified by
    introducing null coordinates in region II, similar to $v$ and $w$
    previously used in region I. Continuations through the past horizons are
    performed in a similar manner.
    The resulting causal structures depend on the parities of $m$ and $n$.

\medskip\noi
    {\bf 1a.} Both $m$ and $n$ are even, so $P(\rho)$ is an even function.
    The two regions $\rho > 0$ and $\rho < 0$ are isometric
    ($g_{\mu\nu}(-\rho) = g_{\mu\nu}(\rho)$), and the Penrose diagram is
    similar to that for the extreme Kerr space-time,
    an infinite tower of alternating regions I and II (\fig 2, left).
    All points of the diagram, except the boundary and the
    horizons, correspond to usual 2-spheres.

\medskip\noi
    {\bf 1b.} Both $m$ and $n$ are odd; again $P(\rho)$ is an even function,
    but regions I and II are
    now anti-isometric ($g_{\mu\nu}(-\rho) = -g_{\mu\nu}(\rho)$).
    The metric tensor in region II ($\rho < 0$) has the signature ($-+++$)
    instead of ($+---$) in region I. Nevertheless, the Lorentzian nature of
    the space-time is preserved. The Penrose diagram is shown in \fig 3,
    left.  In both cases 1a and 1b the maximally extended space-times are
    globally regular\footnote{A globally regular extension of an extreme
    dilatonic black hole, with the same Penrose diagram as in our case 1a,
    was discussed in \cite{GHT}.}.

\medskip\noi
    {\bf 2.} $m-n$ is odd, i.e. $P(\rho)$ is an odd function; moreover,
    $P \to -\infty$ ($r \to 0$) as $\rho \to -\infty$, so that the
    metric (\ref{global}) is singular on the line $\rho = -\infty$, which is
    spacelike. The resulting Penrose
    diagrams are similar to those of the Schwarzschild
    space-time in case 2a ($n$ odd, $m$ even), and of the extreme ($e^2 =
    m^2$) Reissner--Nordstr\"{o}m space-time in case 2b ($n$ even, $m$ odd,
    \fig 4).  In the latter subcase, however, the 4-dimensional metric
    changes its signature when crossing the horizon, similarly to case 1b.

\subsection {Type B2 structure}                           

    Let us briefly consider the case B2: \ $k>0$, $a>\xi>1$. As before,
    the metric is transformed according to (\ref{vw})--(\ref{VW}) and at the
    future null limit (now infinity rather than a horizon, therefore we
    avoid the term ``\bh") we again arrive at (\ref{ext}), where now $W\to
    \infty$ as $P\to 0$. The asymptotic form of the metric as $W\to \infty$
    is
\beq
	ds^2 =  -C_1 dv\,dW - C_2 W^{-m} d\Omega^2      \label{B2vW}
\eeq
    where $C_{1,2}$ are some positive constants, while
    the constant $m$, defined in (\ref{def-m}), is
    now negative.  A further application of the $v$-transformation
    (\ref{VW}) at the same asymptotic, valid for any finite $v<0$, leads to
\bearr                                                     \label{B2VW}
	ds^2 = -C_1 (-V)^{(a-\xi)/(\xi-1)}dV\,dW - C_2 W^{-m}d\Omega^2.
    \nnn
\ear
    If we now introduce new radial ($R$) and time ($T$) coordinates
    by $T=V+W$ and $R=T-W$, in a spacelike section $T=\const$ the limit
    $R\to -\infty$ corresponds to simultaneously $V\to -\infty$ and
    $W\to +\infty$, with $|V|\sim W$, and the metric (\ref{B2VW}) turns into
\bear
	ds^2 \eql 4C_1 (-R)^{(a-\xi)/(\xi-1)}(dT^2 - dR^2) \nnn\label{B2RT}
    \inch                    - C_2 (-R)^{-m}d\Omega^2.
\ear
    Evidently, this asymptotic is a nonflat spatial infinity, with infinitely
    growing coordinate spheres and also infinitely growing $g_{00}$, i.e.,
    this infinity repels test particles.
 \WPicture{79}
 {\input bh3.pic
  \cm
 \input bh4.pic
  }
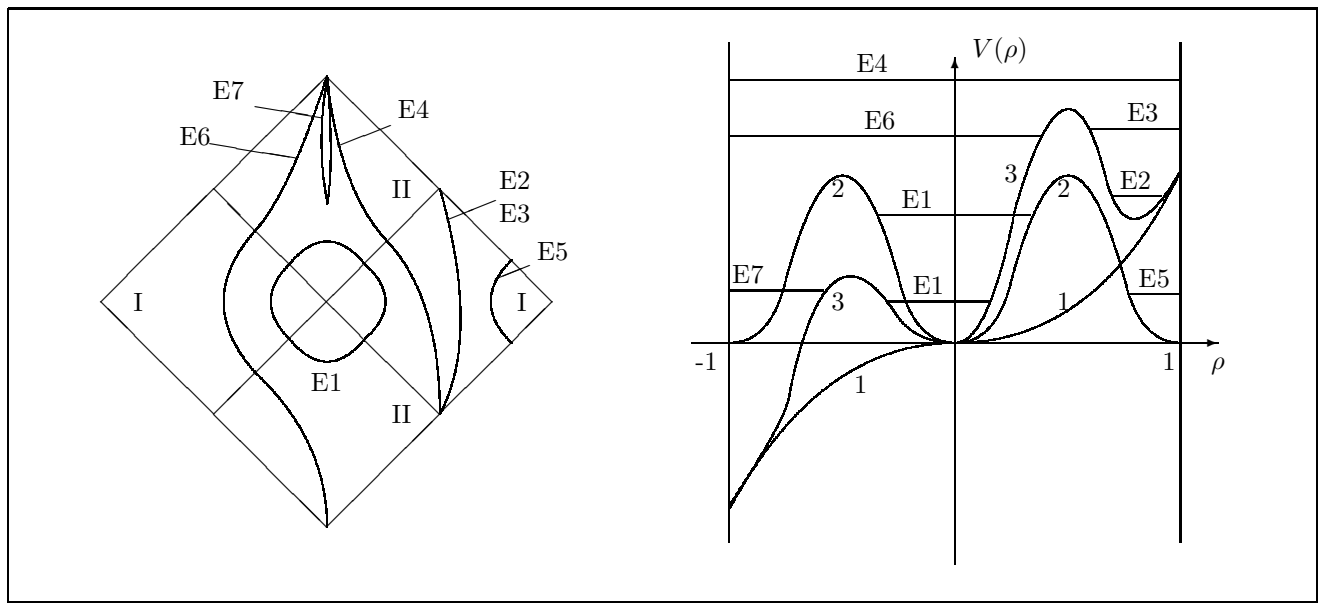
 {The Penrose diagram and the effective potential for geodesics for a BH
 with $m$ and $n$ both odd.  Curve 1 is $V_\eta(\rho)$, curve 2 is
 $V_L(\rho)$ and curve 3 is $V(\rho)$ for a nonradial timelike geodesic. The
 curves and ``energy levels" E1--E7 correspond to different kinds of
 geodesics as described in \ssect 4.4, item 1b.}

    A Penrose diagram of a B2 type configuration coincides with a single
    region I in any of the diagrams; all its sides depict null infinities,
    its right corner corresponds to the usual spatial infinity and its left
    corner to the unusual one, described by the metric (\ref{B2RT}). The
    latter has been obtained here by ``moving along" the future null
    infinity $W\to\infty$, but the same is evidently achieved starting from
    the past side.

\subsection{Geodesics}           

    Let us now return to type B1 BHs and study test particle motion in
    their backgrouds, described by geodesics equations.

    One can verify that all geodesics are continued smoothly across the
    horizons, even in cases 1b and 2b when the metric changes its
    signature (the geodesic equation depends only on the Christoffel symbols
    and is invariant under the anti-isometry $g_{\mu\nu} \to -g_{\mu\nu}$).

    We will use the metric (\ref{global}). Then the integrated geodesic
    equation for arbitrary motion in the plane $\theta=\pi/2$ reads:
\bearr                                                       \label{geo}
\nq     \frac{4k^2(m-n)^2}{(1-P)^4}\dot{\rho}^2
	+ \eta \rho^{n+2}
     + \frac{L^2}{4k^2}(1-P)^2\rho^{m+n+2} = E^2 \nnn
\ear
    where $\dot{\rho} \equiv d\rho/d\lambda$, $\lambda$ being an affine
    parameter such that $ds^2 = \eta d\lambda^2$, with $\eta = +1,\ 0,\ -1$
    for timelike, null or spacelike geodesics; $E^2$ and $L^2$ are constants
    of motion associated with the timelike and azimuthal Killing vectors and,
    correspondingly, with the particle total energy and angular momentum.
    \eq(\ref{geo}) has the form of an energy balance equation,
    with the effective potential
\beq
	 V(\rho)= V_{\eta}(\rho) + V_L(\rho)             \label{V}
\eeq
    where $V_{\eta}$ and $V_L$ are the respective terms in the left-hand
    side.

    One should note that although the coordinate $\rho$ belongs to the
    static frame of reference, one can use it (and consequently \eq
    (\ref{geo})) to describe geodesics that cross the horizon since
    in a close neighbourhood of the horizon $\rho=0$ it coincides (up to a
    positive constant factor) with a manifestly well-behaved coordinate $V$
    or $W$ and, on the other hand, \eq (\ref{geo}) reads simply
    $\dot\rho{}^2 = \const\cdot E^2$ and is thus also well-behaved.

    Let us discuss the four possible cases according to the
    parity of $m$ and $n$ and the corresponding signature of the metric
    (\ref{global}) for $\rho < 0$:

\medskip\noi
{\bf 1a:} $m$ even, $n$ even, ($+ - - \,-$). The range of $\rho$ is
    $(-1,+1)$. The coefficient $(1-P)^{-1}$ of the kinetic term and the
    potential in (\ref{geo}) are both symmetric under the exchange
    $\rho \to -\rho$. The potential $V(\rho)$ is shown in
    Fig.\,2: curve 1 depicts $V_\eta(\rho)$ for $\eta=1$, i.e., the
    potential for radial timelike geodesics; curve 2 shows $V_L$, the
    angular momentum dependent part of $V(\rho)$, and curve 3 shows
    their sum for certain generic values of the motion parameters.
    Depending on the value of $E$, geodesic motion can be symmetrical with
    successive horizon crossings from one region to the next isometrical
    region without reaching past or future null infinity (E1: see ``energy
    levels" in Fig.\,2, right and the corresponding curves in Fig.\,2,
    left), or starting from a past timelike infinity in region I and
    reaching a future timelike infinity in region II (E4).  Some nonradial
    timelike (E2, E3) and null (E5) geodesics remain in a single region,
    corresponding to bound (E2) or unbound (E3) particle orbits near the BH
    or photons passing it by (E5). The existence of nonradial trajectories
    like E1 for any value of $L$ is here connected with  the infinite value
    of $\e^{\beta}$ at the horizon, creating a minimum of $V_L$. Radial null
    geodesics ($V(\rho)\equiv 0$) correspond, as always, to straight lines
    tilted by $45^{\circ}$ (unshown).

\medskip\noi
{\bf 1b:} $m$ odd, $n$ odd, ($- + + \,+$).
    The one-dimensional dependence $\rho(\lambda$) is qualitatively the same
    for null geodesics ($\eta = 0$). However, the global picture is
    drastically different, see \fig 3. In particular, type E1 null
    geodesics which periodically cross the horizon necessarily go repeatedly
    through all four regions of the Penrose diagram, so that their
    projections onto a 2-dimensional plane $\theta= \const$, $\Phi=\const$
    are closed; they will be even closed in the full 4-dimensional
    space-time for some discrete values of the angular momentum $L^2$. Thus
    such BH space-times contain closed null geodesics, leading to
    causality violation.
 \WPicture{81}
  {\input bh6.pic \cm \quad
   \input bh5.pic}
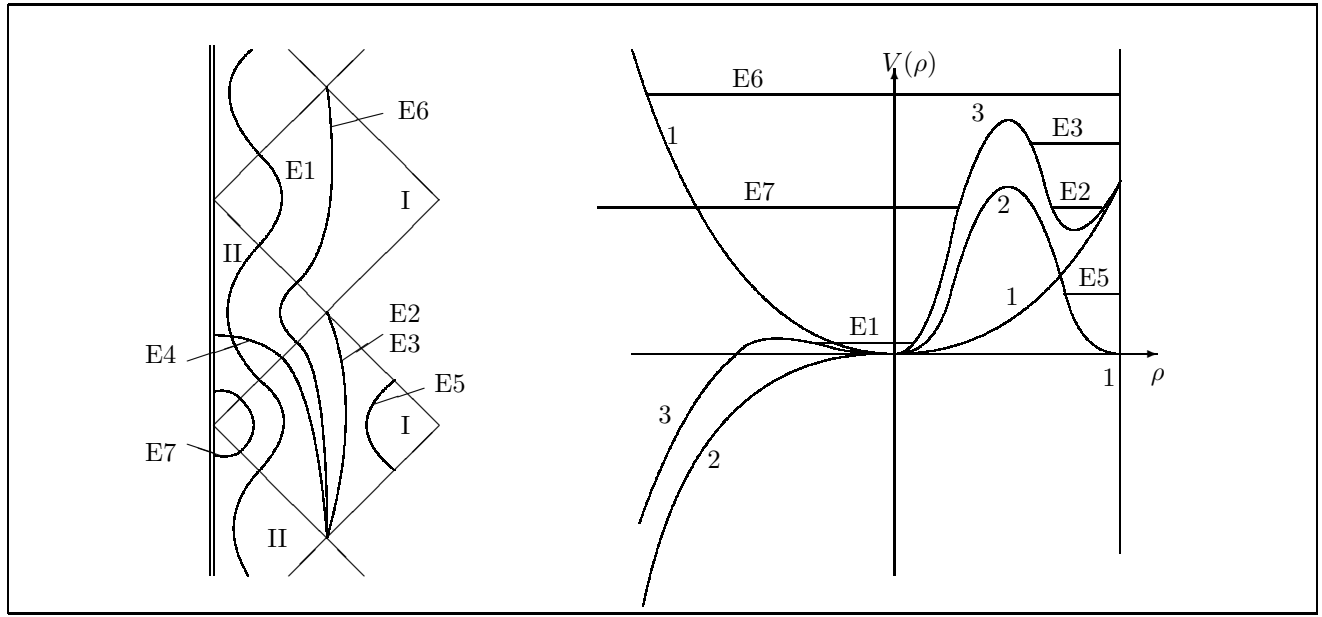
 {The Penrose diagram and the effective potential for geodesics for a BH
 with $n$ even and $m$ odd.  Curve 1 is $V_\eta(\rho)$, curve 2 is
 $V_L(\rho)$ and curve 3 is $V(\rho)$ for a nonradial timelike geodesic.
 The curves and ``energy levels" E1--E7 correspond to different kinds of
 geodesics as described in \ssect 4.4, item 2b.}

    However, this problem can be avoided if we choose, instead of the
    simplest, one-sheet maximal analytic extension corresponding to the
    planar Penrose diagram of \fig 3, left, a ``helicoidal" analytic
    extension constructed in the following way: starting from a given region
    I and proceeding with the extension counterclockwise, after 4 steps we
    come again to a region I isometric to the original one, but do not
    identify these mutually isometric regions and repeat the process
    indefinitely.  The same process is performed in the clockwise direction.
    The resulting Penrose diagram is a Riemann
    surface with a countable infinity of sheets similar to that in \fig 3,
    left, cut along one of the horizons. Such a structure no longer
    exhibits causality violation%
\footnote
{Strictly speaking, such a process might be applied to the Schwarzschild
 and Rindler space-times, with possible identification of isometric regions
 after a finite number of steps. This is, however, unnecessary since there
 a causality problem like ours does not exist.}.
    In general, when crossing a horizon, a null geodesic remains null,
    but timelike geodesics become spacelike and {\sl vice versa\/},
    since the coefficient $\eta$ in \eq (\ref{geo}), being an integral of
    motion for a given geodesic, changes its meaning in transitions between
    regions I and II: a geodesic with $\eta=1$ is timelike in region I and
    spacelike in region II, and conversely for $\eta=-1$.

    For nonnull geodesics the potential is now asymmetric: thus, for
    trajectories which are timelike in region I ($\eta = +1$), it becomes
    attractive for $\rho < 0$ (Fig.\,3, the notations coincide with those of
    Fig.\,2).  These geodesics become spacelike in region II and
    generically extend to spacelike infinity as $\lambda \to \infty$ (E4 in
    Fig.\,3). This is true for all radial geodesics and part of nonradial
    ones; however, nonradial geodesics with small E (near the minimum of
    curve 3 at $\rho=0$) are of the type E1, quite similar to null E1
    trajectories that have been just discussed. A new type of tachyonic
    motion as compared with item 1a is E6 shown in \fig 3.
    There are also circular unstable geodesics with $\eta=+1$ in region II,
    with $t=\const$ and $\rho=\rho_0 =-[m/(3m-2n)]^{1/(m-n)}$
    (points in the Penrose diagram), such that $V(\rho_0) = V'(\rho_0) = 0$,
    $E=0$. All these spacelike geodesics have full analogues with $\eta=-1$
    in region I.

    The whole space-time possesses full symmetry under an exchange
    between regions I and II, corresponding to rotations of the Penrose
    diagram of \fig 3 by odd (in addition to even) multiples of the right
    angle, accompanied by a change of sign of $\eta$ so that geodesics
    keep their timelike or spacelike nature.

    \medskip\noi
{\bf 2a:} $m$ even, $n$ odd, $(- + - \,-)$. The range of $\rho$ is now
    $(-\infty, +1)$. Both parts of the effective potential become
    negative and monotonic at $\rho < 0$, so that all geodesics
    entering the horizon terminate at the spacelike singularity $\rho =
    -\infty$, as in the Schwarzschild case. Thus the whole
    qualitative picture of test particle motion, as well as the Penrose
    diagram, coincide with the Schwarzschild ones.

\medskip\noi
{\bf 2b:} $m$ odd, $n$ even, $(+ - + \,+)$. The
    potential $V_\eta$ ($\eta=1$) is positive-definite (as in the extreme
    Reissner--Nordstr\"{o}m case), as shown in Fig.\,4, so that all radial
    timelike geodesics avoid the singularity,
    crossing the horizon either twice (E6),
    or indefinitely (E1). For
    nonradial motion, the summed potential $V(\rho)\to -\infty$ as
    $\rho\to -\infty$, whatever small is $L^2$, therefore all geodesics that
    are timelike in region I (and become spacelike in region II) reach the
    spacelike singularity%
\footnote
{Due to the metric signature change, in the Penrose diagram of Fig.\,4, just
as in Fig.\,3 for the case 1b, the time direction is vertical in regions I
and horizontal in regions II.}
    $\rho=-\infty$ (levels and curves E4 and E7 in Fig.\,4), except
    those with small $E$ depicted as E1.
\medskip

    One can conclude that the unusual nature of metric of B1 type BHs
    creates some unusual types of particle motion. Some of them even exhibit
    evident causality violation
    --- such as an observer receiving messages from his or her own future ---
    which can be avoided by assuming a more complicated space-time structure.

    Another paradox, also related to causality, arises if we
    follow in cases 1b or 2b the fate of a hypothetical (timelike) observer
    who has crossed the horizon and finds him(her)self in a region II where
    his (her) proper time is now spacelike as viewed by a resident
    observer (whose timelike geodesic is entirely contained in region II),
    and can be reversed by a simple coordinate change.

    However, one may suspect that, the horizon area
    being infinite, any extended body, and an observer in particular,
    will have been infinitely stretched apart
    and destroyed before actually crossing the horizon. To check this,
    consider for instance a freely falling observer whose center of mass
    follows a radial geodesic in the plane $\theta = \pi/2$. The separation
    $n^{\alpha}$ between this geodesic and a neighbouring radial geodesic
    varies according to the law of geodesic deviation
\beq                                                     \label{dev-geo}
    \frac{D^2n^{\alpha}}{d\lambda^2} +
    {R^{\alpha}}_{\beta\gamma\delta}u^{\beta}n^{\gamma}u^{\delta} = 0.
\eeq
    For the four--velocity of the center of mass we have
    $u^0 = E g^{00}$ and $u^0 u_0 + u^1 u_1=1$,
    so that near the horizon
    $u^{\mu} \simeq (Eg^{00}, E(-g^{00}g^{11})^{1/2}, 0, 0)$. We obtain for
    the relative azimuthal acceleration near the horizon
\bearr                                                    \label{tide}
	\frac{1}{n^3}\frac{D^2 n^3}{d\lambda^2}
	= ({R^{30}}_{30}u^0 u_0 + {R^{31}}_{31}u^1 u_1)
\nnn \qquad
	\simeq 	-E^2 g^{00}({R^{30}}_{30} - {R^{31}}_{31})
	= E^2 R''/R,
\ear
    and a similar equation for the deviation $n^2$ in the $\theta$
    direction. Here $R^2 =|g_{22}|$, $R'' = d^2 R/d\rho^2$, and $\rho$ is a
    radial coordinate
    such that the Jordan--frame metric functions are related by
    $g_{00}g_{11} = \const$; this condition is valid near the horizon for
    our coordinate $\rho$ defined in (\ref{rho}).

    The azimuthal geodesic deviation (\ref{tide}) which, due to its
    structure, is insensitive to radial boosts and is thus equally applicable
    to the static frame of reference and to the one comoving with the
    infalling body, agrees with similar relations given by Horowitz and Ross
    \cite{HR}.  In the case of the Schwarzschild metric, $R''/R = 0$ and the
    tidal force (given by the terms that we have neglected) is finite. In
    the case of the examples discussed in \cite{HR}, $R''/R$ is negative and
    large, i.e. geodesics converge and physical bodies are crushed as they
    approach the horizon, as by a naked singularity, hence the name ``naked
    black holes'' given to these spacetimes in \cite{HR}. On the contrary,
    in the case of the cold black hole metric (\ref{global}), $R''/R \to
    +\infty$ (as $\rho^{-2}$), i.e. geodesics diverge; the resulting
    infinite tidal forces pull apart all extended objects, e.g. any kind of
    clock, approaching the horizon. Only true elementary (pointlike)
    particles, resulting from the destruction of the falling body, cross
    such a horizon to become tachyons\footnote
{Our cold black holes are thus
counterexamples to the claim made in \cite{HR} that a smooth extension
is not possible when tidal forces diverge on the horizon.}.


\section{Stability}

    Let us now study small (linear) \sph \pns\ of the above static
    solutions (or static regions of the BHs), i.e. consider, instead of
    $\varphi(u)$,
\beq
	\varphi(u,t)= \varphi(u)+ \df(u,t)
\eeq
    and similarly for the metric functions $\alpha,\beta,\gamma$, where
    $\varphi (u)$, etc., are taken from the static solutions of \sect 2.
    We are working in the Einstein conformal frame and use \eqs (\ref{eq1}).
    The consideration applies to the whole class of STT (\ref{L1}); its
    different members can differ in boundary conditions, to be
    discussed below.

    In perturbation analysis there is the so-called gauge freedom, i.e. that
    of choosing the frame of reference and the coordinates of the perturbed
    space-time.  The most frequently used frame for studying radial
    perturbations has been that characterized by the technically convenient
    condition $\db \equiv 0$ \cite{78,hod,bm}.  It was applied, however, to
    background configurations where the area function $\e^\beta$ was
    monotonic in the whole range of $u$, or was itself used as a coordinate
    in the static space-time.  Unlike that, in our study the configurations
    of utmost interest are type B \bhs\ and wormholes, i.e. those having a
    minimum of $\e^\beta$ (a throat) at some value of $u$ and infinite
    $\e^{\beta}$ at both ends of the $u$ range.  At the throats, the
    equality $\db\equiv 0$ is no longer a coordinate condition, but a
    physical restriction, forcing the throat to be at rest. It can be
    explicitly shown that the condition $\db\equiv 0$ creates poles in the
    effective potential for \pns, leading to their separate existence at
    different sides of the throat, i.e. the latter behaves like a wall.

    For these reasons, we have to use another gauge,
    and we choose it in the form
\beq
	\da = 2\db + \dg,                                     \label{da}
\eeq
    extending to \pns\ the harmonic coordinate condition of the static
    system.  In this and only in this case the scalar equation (\ref{ephi})
    for $\df$ decouples from the other \pn\ equations and reads
\beq
	\e^{4\beta(u)}\delta\ddot\varphi  - \df''=0.          \label{edf}
\eeq
    Since the scalar field is the only dynamical degree of freedom, this
    equation can be used as the master one, while others only express the
    metric variables in terms of $\df$, provided the whole set of field
    equations is consistent. That it is indeed the case, can be verified
    directly. Indeed, under the condition (\ref{da}), the four equations
    (\ref{00})--(\ref{01}) for \pns\ take the form
\bearr                                                \label{00p}
    \e^{4\beta}(4\db +\dg)\ddot{} - \dg'' =0; \\ \lal
                                                      \label{11p}
    \e^{4\beta}(2\db +\dg)\ddot{} -2\db'' -\dg''  \nnn
    \cm
	  -4(\beta'-\gamma')\db'+4\beta'\dg' = 2\eps\varphi'\df'; \\ \lal
										    \label{22p}
    \e^{4\beta}\delta\ddot{\beta}
	  -\db'' + 2 \e^{2\beta+2\gamma}(\db+\dg) =0; \\ \lal
										    \label{01p}
    \dot{\beta}' - \beta'(\db + \dg)\dot{}
	                  - \gamma'\delta\dot{\beta}
		= -\half \eps \varphi'\delta\dot{\varphi},
\ear
    where $\alpha,\ \beta,\ \gamma,\ \varphi$ satisfy the static field
    equations.

    \eq(\ref{01p}) may be integrated in $t$ and further
    differentiated in $u$; the result turns out to be proportional to a
    linear combination of the remaining Einstein equations
    (\ref{00p})--(\ref{22p}). On the other
    hand, the quantities $\delta\ddot\varphi$ and $\df''$ can be calculated
    from (\ref{00p})--(\ref{01p}), resulting in (\ref{edf}). Therefore we
    have three independent equations for the three functions $\df$, $\db$
    and $\dg$.

    The following stability analysis rests on \eq (\ref{edf}).
    The static nature of the background solution makes it possible to
    separate the variables,
\beq
	\df = \psi(u) \e^{i\omega t},                    \label{psi}
\eeq
    and to reduce the stability problem to a boundary-value problem for
    $\psi(u)$. Namely, if there exists a nontrivial solution to (\ref{edf})
    with $\omega^2 <0$, satisfying some physically reasonable boundary
    conditions at the ends of the range of $u$, then the static background
    system is unstable since \pns\ can exponentially grow with $t$.
    Otherwise it is stable in the linear approximation.

    Suppose $-\omega^2 = \Omega^2,\ \Omega > 0$. In what follows we
    use two forms of the radial equation (\ref{edf}): the one directly
    following from (\ref{psi}),
\beq
	\psi'' -\Omega^2 \e^{4\beta(u)}\psi=0,             \label{epsi}
\eeq
    and the normal Liouville (Schr\"odinger-like) form
\bearr
	d^2 y/dx^2 - [\Omega^2+V(x)] y(x) =0,   \nnn \cm
	V(x) = \e^{-4\beta}(\beta''-\beta'{}^2).            \label{ey}
\ear
    obtained from (\ref{epsi}) by the transformation
\beq
	\psi(u) = y(x)\e^{-\beta},\qquad                     \label{tx}
				x = - \int \e^{2\beta(u)}du.
\eeq
    Here, as before, a prime denotes $\d/\d u$.

    The boundary condition at spatial infinity ($u\to 0$, $x \simeq 1/u \to
    +\infty$) is evident: $\df\to 0$, or $\psi\to 0$.
    For our metric (\ref{ds1}) the effective potential $V(x)$ has the
    asymptotic form
\beq
    V(x) \approx 2b/x^3, \cm {\rm as} \cm x\to +\infty,
\eeq
    hence the general solutions to (\ref{ey}) and (\ref{epsi}) have the
    asymptotic form
\bear
    y \al\sim \al c_1\e^{\Omega x} + c_2\e^{-\Omega x}
    \qquad (x \rightarrow +\infty),            \label{as+} \\
    \psi \al                                                \label{aspsy}
        \sim \al u \bigl(c_1 \e^{\Omega/u} + c_2\e^{-\Omega/u}\bigr)
    \qquad (u \rightarrow 0),
\ear
    with arbitrary constants $c_1,\ c_2$. Our boundary condition leads to
    $c_1=0$.

    For $u\to \umx$, where in many cases the background field $\varphi$
    tends to infinity, a formulation of the boundary condition is not so
    evident. Refs.\,\cite{hod,bm} and others, dealing with
    minimally coupled or dilatonic scalar fields, used the minimal
    requirement providing the validity of the \pn\ scheme in the Einstein
    frame:
\beq
    |\df/\varphi| < \infty.                                  \label{weak}
\eeq
    In STT, where Jordan-frame and Einstein-frame metrics are related by
    $g^{\rm J}_{\mu\nu} = (1/\phi)g^{\rm E}_{\mu\nu}$,
    it seems reasonable to require that the perturbed
    conformal factor $1/\phi$ behave no worse than the unperturbed one,
    i.e.
\beq
    |\delta\phi/\phi| < \infty.                         \label{strong}
\eeq
    An explicit form of this requirement depends on the specific STT and
    can differ from (\ref{weak}), for example, in the BD theory, where
    $\phi$ and $\varphi$ are connected by (\ref{phi-bd}), the requirement
    (\ref{strong}) leads to $|\df| <\infty$. We will refer to (\ref{weak})
    and (\ref{strong}) as to the ``weak" and ``strong" boundary condition,
    respectively. For configurations with regular $\phi$ and $\varphi$ at
    $u\to \umx$ these conditions both give $|\df|<\infty$.

    Let us now discuss different cases of the STT solutions.
    We will suppose that the scalar field $\phi$ is regular for
    $0<u<\umx$, so that the conformal factor $\phi^{-1}$ in (\ref{ds1})
    does not affect the range of the $u$ coordinate.

\medskip\noi
    {\bf 1.} $\eps=+1,\ k>0$. This is the singular solution of normal STT.
    As $u \to +\infty$, $\beta \sim (b-k)u \to -\infty$, so that $x$ tends to
    a finite limit and with no loss of
    generality one can put $x\to 0$. The effective potential $V(x)$ then has a
    negative double pole, $V \sim -1/(4x^2)$, and the asymptotic form of the
    general solution to (\ref{ey}) leads to
\beq
    \psi(u) \approx y(x)/\sqrt{x} \approx (c_3 + c_4\ln x) \quad
    (x \to 0).             \label{y1}
\eeq

    The weak boundary condition leads to the requirement
    $|\delta\varphi/\varphi| \approx
    |y|/(\sqrt{x}|\ln x|) < \infty$, met by the general solution (\ref{y1})
    and consequently by its special solution that joins the allowed
    case ($c_1=0$) of the solution (\ref{as+}) at the spatial asymptotic.
    We then conclude that the static field configuration is unstable,
    in agreement with the previous work \cite{hod}.

    As for the strong boundary condition (\ref{strong}),
    probably more appropriate in STT, its explicit form
    varies from theory to theory, therefore a general conclusion cannot be
    made. In the special case of the BD theory the condition (\ref{strong})
    means $|\psi|< \infty$ as $u\to +\infty$. Such an asymptotic behavior
    is forbidden by (\eq (\ref{epsi}), according to which $\psi''/\psi > 0$,
    i.e. the perturbation $\psi(u)$ is convex and so cannot be bounded as $u
    \to \infty$ for an initial value $\psi(0) = 0$ ($c_1 = 0$). We thus
    conclude that the static system is stable.

    We see that in this singular case the particular choice of a boundary
    condition is crucial for the stability conclusion. In the case of
    GR with a minimally coupled scalar field \cite{hod}
    there is no reason to ``strengthen" the weak condition that leads to
    the instability. In the BD case the strong condition seems more
    reasonable, so that the BD singular solution seems stable. For any
    other STT the situation must be considered separately.

\medskip\noi
    {\bf 2.} $\eps=-1,\ k>0$. This case includes some singular
    solutions and cold \bhs, as is exemplified for the BD theory in
    (\ref{ds+})--(\ref{range+}).

    As $u\to +\infty$, $\beta \sim (b-k)u \to +\infty$, so that $x\to
    -\infty$ and $V(x) \approx -1/(4x^2)\to 0$.  The general solution to \eq
    (\ref{ey}) again has the asymptotic form (\ref{as+}) for $x \to
    -\infty$.  The weak condition (\ref{weak}) leads, as in the previous
    case, to the requirement $|y|/(\sqrt{|x|}\ln|x|) <\infty$, and, applied
    to (\ref{as+}), to $c_2=0$. This means that the function $\psi$ must
    tend to zero for both $u\to 0$ and $u\to \infty$, which is impossible
    due to $\psi''/\psi >0$. Thus the static system is stable. Obviously the
    more restrictive strong condition (\ref{strong}) can only lead to the
    same conclusion.

\medskip\noi
    {\bf 3.} $\eps=-1,\ k=0$. There are again singular solutions and cold
    \bhs. As $u \to +\infty$, $x\to -\infty$ and again the potential $V(x)
    \approx -1/4x^2 \to 0$, leading to the same conclusion as in case 2.

\medskip\noi
    {\bf 4.} $\eps=-1,\ k<0$. In the generic case the solution describes a
    wormhole, and in the exceptional case (\ref{bh-}) there is a cold
    \bh\ with a finite horizon area. In all such cases, as $u\to\umx =
    \pi/|k|$, one has $x\to -\infty$ and $V \sim 1/|x|^3 \to 0$, so that the
    stability is concluded just as in cases 2 and 3.

    Thus, generically, scalar-vacuum \sph solution of anomalous STT are
    linearly stable against \sph \pns.  Excluded are only the cases when the
    field $\phi$ behaves in a singular way inside the coordinate range $0 <u
    <\umx$; such cases should be studied individually.

\Acknow
{This work was supported in part by CNPq (Brazil) and CAPES (Brazil).}

\end{document}